\documentclass[prl,reprint]{revtex4-1}
\usepackage{graphicx}
\usepackage{dcolumn}
\usepackage{bm}
\usepackage{graphicx}
\usepackage{dcolumn}
\usepackage{bm}
\usepackage{subfigure}
\usepackage{color}
\usepackage{amsmath}

\begin{document}


\title{Reynolds number of transition and large-scale  properties of  strong turbulence.
} 


\author{Victor Yakhot}
\affiliation{Department of Mechanical Engineering, Boston University, Boston, Massachusetts 02215, USA.}



\date{\today}

\begin{abstract}  

\noindent  A turbulent flow is characterized by velocity fluctuations excited in  an extremely 
 broad interval  of wave numbers  $k> \Lambda_{f}$ where $\Lambda_{f}$ is a relatively small 
  set of the wave-vectors 
 where energy is pumped into  fluid  by external forces.  
 Iterative averaging over small-scale velocity fluctuations  from the interval  $\Lambda_{f}< k\leq \Lambda_{0}$,
where  $\eta=2\pi/\Lambda_{0}$ is the dissipation scale,    leads to  an infinite number of ``relevant'' scale-dependent coupling constants ( Reynolds numbers ) $Re_{n}(k)=O(1)$.  
It is shown that  in the i.r. limit $k\rightarrow \Lambda_{f}$,  the Reynolds numbers $Re(k)\rightarrow Re_{tr}$ where $Re_{tr}$ is  the recently numerically and experimentally discovered universal Reynolds number of  ``smooth'' transition from Gaussian to anomalous statistics  of spatial velocity derivatives. 
The calculated  relation $Re(\Lambda_{f})=Re_{tr}$  ``selects'' the lowest - order non-linearity  as the only relevant one.  
  This means that in the infra-red limit $k\rightarrow \Lambda_{f}$ all high-order nonlinearities generated by the scale-elimination sum  up to zero.
 \end{abstract}

 \maketitle
PACS numbers 47.27\\

{\bf  Introduction.}   ``The turbulence problem''  can be formulated as follows: consider the Navier-Stokes (NS) equations driven  by the large-scale  force ${\bf F}(\Lambda_{f})$ where $\Lambda_{f}$ denotes a  relatively small set of wave-vectors $|{\bf k}|\approx  2\pi /L$.  We fix  both  force 
 $F=O(1)$ and the integral scale $L=2\pi/\Lambda_{f}=O(1)$ independent upon Reynolds number, and by decreasing kinematic viscosity $\nu$,  vary the Reynolds number $Re=uL/\nu$, where ${\bf u}({\bf r},t)$ is a solution to the NS equations of motion.    As long as $\nu>\nu_{tr}$  ($Re<Re_{tr})$ the flow is {\bf laminar}, i.e.  ${\bf u=u(k)}$ with $k\approx \Lambda_{f}$.  
At $Re=Re_{tr}$ ($\nu=\nu_{tr}$) the solution $u=u_{0}(\Lambda_{f})$ becomes unstable and at $Re>Re_{tr}$, the velocity field can be written as: ${\bf u}({\bf k})={\bf u_{0}}(\Lambda_{f})+{\bf v}({\bf k},t)$ where $k>\Lambda_{f}$.  Formation of the small-scale time-dependent velocity components ${\bf v}({\bf k},t)$ is the main manifestation of transition to turbulence.  {\it We would like to stress a relatively trivial,  but extremely important for  what follows,  statement: at $Re\leq Re_{tr}$ the  laminar flow pattern ${\bf u}_{0}(\Lambda_{f})$ is a solution to the Navier-Stokes equations characterized by a single coupling constant which is a properly chosen Reynolds number.}
 
 \noindent When the Reynolds number $Re\gg Re_{tr}$ the flow is characterized by velocity fluctuations  ${\bf v}(k)$  excited in a broad interval of scales  $\Lambda_{f}\leq k  \leq \Lambda_{0}$,  and in the limit $Re\rightarrow \infty$ the ratio $\Lambda_{f}/\Lambda_{0}\rightarrow 0$. Below we set the u.v. cut-off $\Lambda_{0}$ equal to Kolmogorov's dissipation scale $\Lambda_{0}\approx \Lambda_{f}e^{\frac{3}{4}}$. 
  In this case, as will be shown below,  the equation for turbulent fluctuations is similar to the Navier-Stokes equation with the broad-band "force" $f_{j}=-v_{i}\partial_{i}u_{0,j}$  in the right side.   Following K.G. Wilson [1] we can average the governing NS equations over  velocity fluctuations from a thin "slice" in the wave-vector space $\Lambda(r)=\Lambda_{0} e^{-r}\leq k \leq \Lambda_{0}$ with  $r\rightarrow 0$. In this case this exact  procedure 
 leads to  equations for the remaining long wave- length modes ${\bf v}(k)$ from the interval $k<\Lambda_{0} e^{-r}$. The main problem is that the derived equation, in addition to corrections to viscosity ($\nu(r)\rightarrow \nu+\Delta \nu(r)$ )  and driving force $\Delta {\bf f}$,  includes an infinite number of coupling constants  (see below) which,  unlike in the theory of critical phenomena, are  relevant  when $r\gg 1$.   Since $\Delta \nu(r)>0$, the $r$-dependent Reynolds number $Re(\Lambda(r))<Re(\Lambda_{0})$. 
 Iterating this procedure one can derive equations for the modes with $k\ll \Lambda_{0} $ belonging to the so called inertial range where molecular ("bare") viscosity $\nu$ is irrelevant and  
 all  coupling constants can depend on the "dressed" Reynolds number $Re(r)=\frac{2\pi v_{rms}(r)}{\nu(r)\Lambda(r)}$. 
 By dimensional reasoning $\nu(r)\approx v_{rms}(r) /\Lambda(r)$ and we see that in the inertial range $Re(r)=O(1)$. This means that the resulting equations include  infinite number of non-linearities generated by the scale-eliminating procedure.  In the inertial range all these terms are relevant and are responsible for  anomalous scaling and non-analyticity of velocity increments . 
 
 \noindent  
  According to the picture described above, the integral scale 
 $L=2\pi/\Lambda_{f}$ is the largest scale of turbulence, i.e. $\Lambda(r)\geq \Lambda_{f}$.  It will be shown below that as $\Lambda(r)\rightarrow \Lambda_{f}$ the effective viscosity $\nu(\Lambda(r))\rightarrow \nu_{tr}$ and $Re(\Lambda(r))\rightarrow Re_{tr}$. Since at this Reynolds number  the  marginally stable  flow,  described by the Navier-Stokes equations with $\nu=\nu_{tr}$ is laminar, we may  conclude that all high-order terms, additional to the NS equations,  must disappear and  only the lowest order  quadratic non-linearity, characteristic of the NS equations survives.   This way transition to turbulence "selects" the only relevant non-linearity. Assuming validity of  Landau's  theory of transition  we will be able conclude that in the limit $\Lambda(r)\rightarrow \Lambda_{f}$, high-order non-linearities generated by the scale-elimination procedure are $O(\sqrt{Re(\Lambda(r))-Re_{tr}})\rightarrow 0$.

 \noindent {\bf  Transition to turbulence.}    There exist  a huge literature on this topic  which,  together with the theory of dynamical systems, evolved into a separate field of research.  Typically, one searches for instabilities  in  laminar flow  ${\bf u}_{0}$  manifested by   exponential growth  of perturbations ${\bf u(k,}t)$. 
{\it We will loosely  identify   laminar flow as  a pattern ${\bf u}_{0}$  formed by a small set of excited modes  supported in the range of wave-numbers $k\approx  \Lambda_{f}$.
 All modes with $k>\Lambda_{f}$ are strongly overdamped, i.e. $u(k)=0$  for both $k\ll \Lambda_{f}$ and $k\gg \Lambda_{f}$. }\\
\noindent  {\it Landau's theory.}  Here we mention just one work which is relevant for  considerations presented below [2]. Assuming that in the vicinity of a transition point imaginary part of complex frequency is much smaller than the real one,  Landau considered the linearized Navier-Stokes equations for incompressible fluid.
Denoting the velocity field at a transition point   ${\bf u}_{0}$ and introducing an infinitesimal perturbation ${\bf u}_{1}$ he wrote  ${\bf u}={\bf u}_{0}+{\bf u}_{1}$ with ${\bf u}_{1}=A(t)f({\bf r})$. Based  on general qualitative considerations,  Landau  proposed:

\begin{eqnarray}
\frac{d|A|^{2}}{dt}=2\gamma|A|^{2}-\alpha |A|^{4}\nonumber
\end{eqnarray}

\noindent where in the vicinity of transition point $\gamma=c(Re-Re_{tr})$ and $\alpha>0$.   In principle, $|A|^{2}$ must be considered as time- averaged.  Landau noted, however, that ${\bf u}_{1}({\bf k})$ is a slow mode and, since the averaging is taken over relatively short time -intervals, the averaging  sign in the above equations  is not necessary.  
At small times the solution exponentially grows and then reaches the maximum 
 $A_{max}\propto \sqrt{Re-Re_{tr}}$.  When  $\gamma=Re-Re_{tr}<0$, any initial perturbation decays.  In  this theory, 
 the magnitude of transitional Reynolds number is a free parameter and since the  large-scale field ${\bf u}_{0}$ strongly depends on geometry, external forces and stresses,   
 the transition Reynolds number  $Re_{tr}$ is not expected to be a universal constant.\\
  \noindent  Landau assumed that  further increase of the Reynolds number leads to instability of first unstable mode generating next two excited modes  with the wave-vectors $k_{2}>k_{1}$ etc.  In  modern lingo, this process can be perceived as an onset  of the energy cascade toward small- scale excitations with $k>\Lambda_{f}$.  
This leads  to formation of ``inertial range'' and  strongly intermittent  small-scale dissipation rate ${\cal E}$. \\

\noindent  {\it Transition to turbulence: a new angle.}  A new way  of looking at  phenomenon of transition to turbulence was  introduced  in numerical simulations  of a flow   at a relatively low Reynolds number $R_{\lambda}=\sqrt{\frac{5}{3{\cal E}\nu}}u_{rms}^{2}\geq 4.0$ [3]. 
{\bf In this approach transition to turbulence is identified with  the first appearance of non-gaussian anomalous  fluctuations of velocity derivatives including those of the dissipation rate ${\cal E}$. As will be shown below, in a sense, it is a transition between two different states :  Gaussian ( structureless ) and  anomalous (structured ), resembling   those observed in experiments on Benard convection.  On a first glance the  transition is smooth meaning that no ``jumps'' in velocity field were detected. However, the precise nature of this transition is yet to be investigated.  All we can state at this point is that the transformation happens in a narrow range of the Reynolds number variation.}
The homogeneous and isotropic turbulence (HIT)   was generated in a periodic box by a force in the right-side of the Navier-Stokes equation
${\bf F}({\bf k},t)={\cal P}\frac{{\bf u}({\bf k},t)}{\sum' |\bf {u}({\bf k},t)|^{2}}\delta_{\bf k,k'}
$, where summation is carried over ${\bf k}_{f}=(1,1,2); \  (1,2,2)$. It is easy to see that the model with this forcing   generates flows with constant energy flux ${\cal P}={\cal E}=\overline{\nu(\frac{\partial u_{i}}{\partial x_{j}})^{2}}=const$ and the variation of the Reynolds number is achieved by  
variation of viscosity.  

\noindent    The results of Ref.[3] can be briefly summarized as  follows: 1. \ Extremely well-resolved simulations of the low-Reynolds number  flows   at $R_{\lambda}\geq 9-10$ revealed a clear scaling range  $M_{n}=\overline{(\frac{\partial u}{\partial x})^{n}} / \overline{(\frac{\partial u}{\partial x})^{2}}^{\frac{n}{2}}
\propto Re^{\rho_{n}}$ 
with  anomalous scaling exponents $\rho_{n}$ consistent with the inertial range exponents typically observed only in  very high Reynolds number flows $Re\gg Re_{tr}$.  Identical scaling exponents $\rho_{n}$  were later obtained in isotropic turbulence generated by a different forcing [4], channel and  pipe  flows [5]  and, more recently,  in Benard convection [6] indicating possibility of a broad universality. 
2. \   For $R_{\lambda}<9-10$ all flows were   subgaussian indicating a dynamical system 
consisting of a small number of modes with  the small-scale fluctuations strongly  overdamped. 
 This flow  can be called ``quasilaminar'' or coherent.   3. \ At  a transition point $R_{\lambda,tr}\approx 9-10$ the fluctuating  velocity derivatives obey gaussian statistics and at $R_{\lambda}>9- 10$ a strongly anomalous scaling of the moments, typical of high-Reynolds number turbulence,  is clearly seen.   4. \ It has also been noticed that transition was smooth, i.e.  velocity field at ${\bf u}(R_{\lambda,tr}^{-})-{\bf u}(R_{\lambda,tr}^{+})\rightarrow 0$.  \\
 \begin{figure}[h]
\center
\includegraphics[height=8cm]{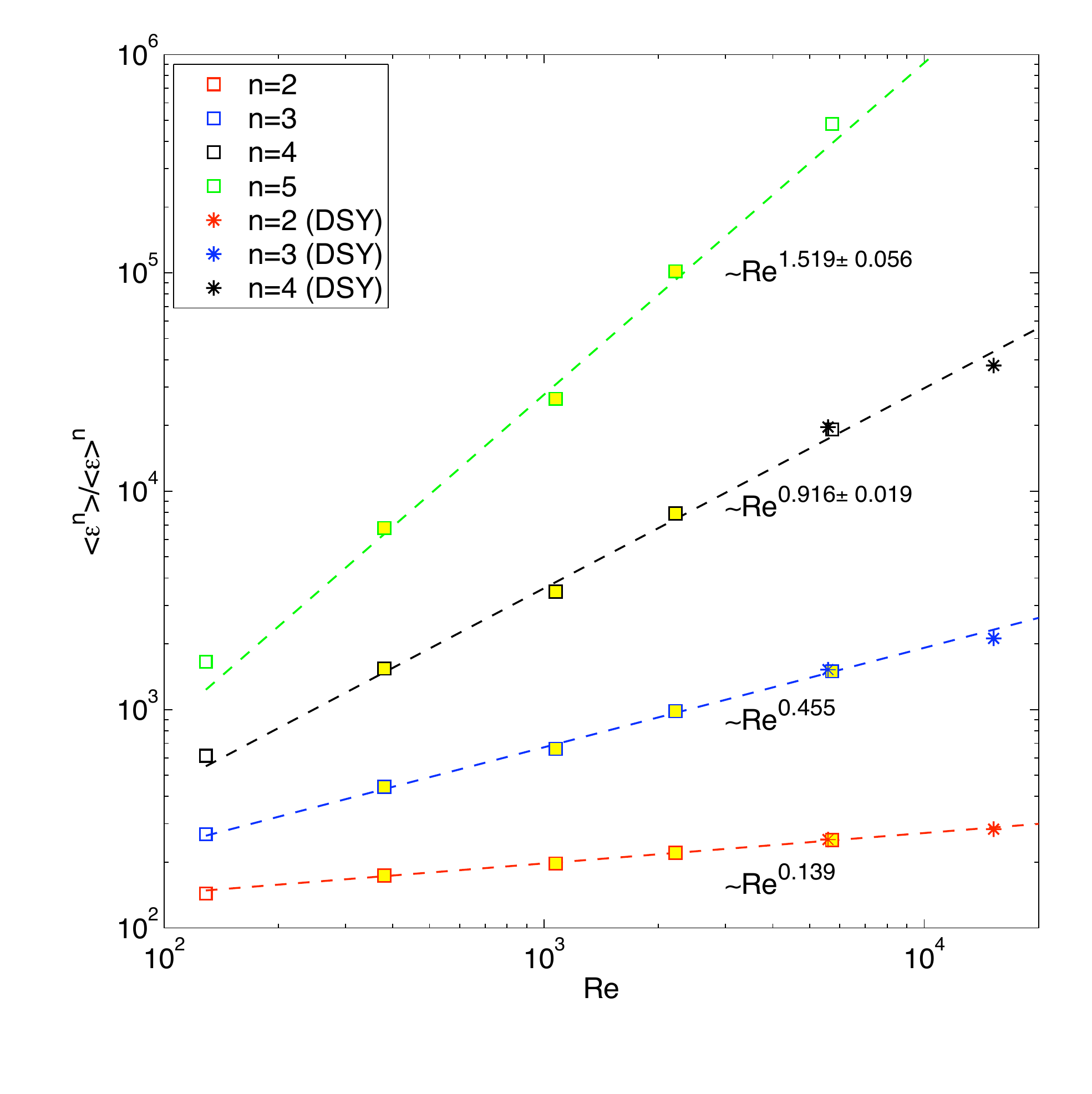}
\caption{(Color online) Normalized moments of the dissipation rate $\frac{\overline{{\cal E}^{n}}}{\overline{{\cal E}}^{n}}$ in homogeneous and isotropic turbulence. Refs.[3]-[4].DSY stand for Donzis, Sreenivasan and Yeung. }\label{figure 1}
\end{figure}

On Fig.1  the moments of the dissipation $\frac{\overline{{\cal E}^{n}}}{\overline{{\cal E}}^{n}}$ rate computed in [3]  
are combined with the data obtained by Donzis et. al [4] in HIT generated by a completely different large-scale forcing.  We can see that the scaling exponents, found in the range of very low Reynolds number in [3]  hold in a much wider  range  of the Reynolds number variation.   This means that in the range $R_{\lambda}\geq 10$ turbulence can be considered as fully developed.  \\
 
\begin{figure}[h!]
	\begin{center}
	\includegraphics[width=.5\textwidth]{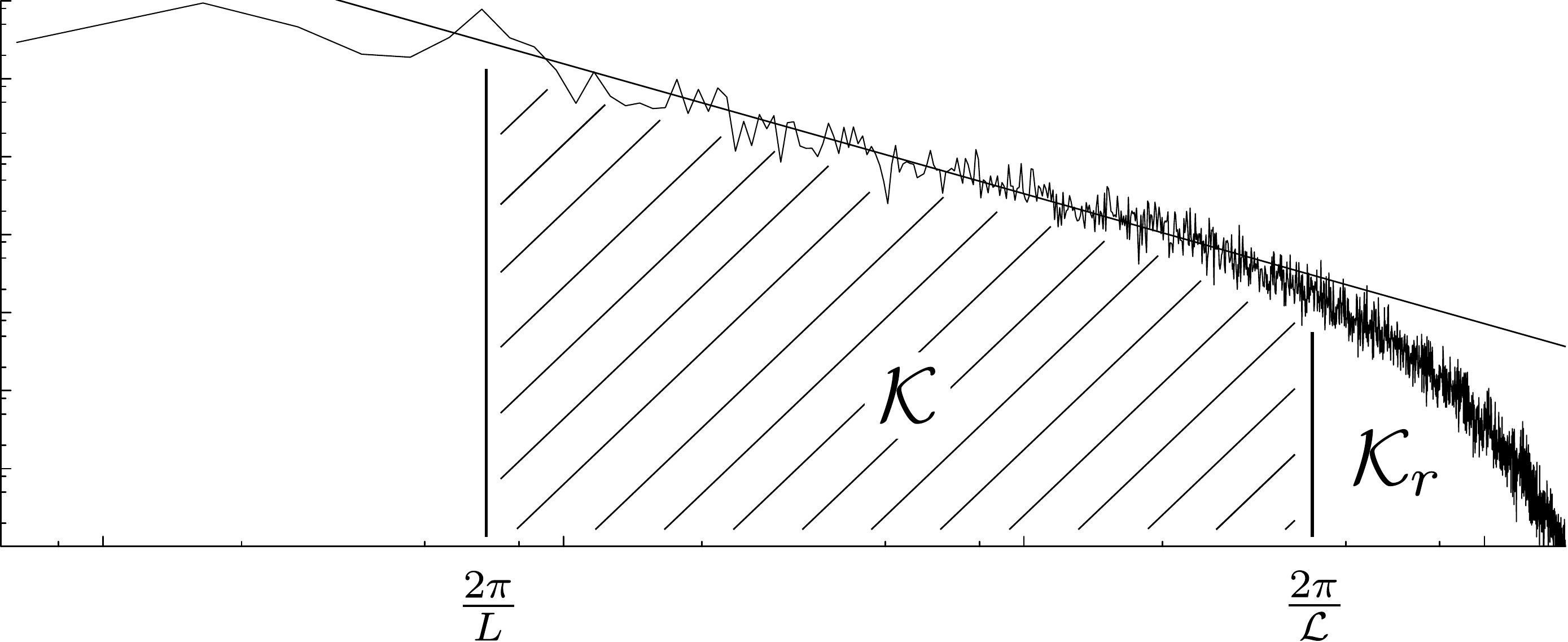}
	\vspace{-0.025\textwidth}
	\end{center}
	\caption{Length-scales and turbulent kinetic energies.  ${\cal K}$   and ${\cal K}_{r}$  denote energy contents in inertial and dissipation ranges, respectively. $L$ and ${\cal L}\approx \eta$ stand for integral scale, corresponding to the top of inertial range and the dissipation scale, respectively. Strait line: $E(k)\approx 1.6 {\cal E}^{\frac{2}{3}}k^{-\frac{5}{3}}$.  }
	\label{fig:scalingRegimes}
\end{figure}

\begin{figure*}
\centering
\includegraphics[height=10cm]{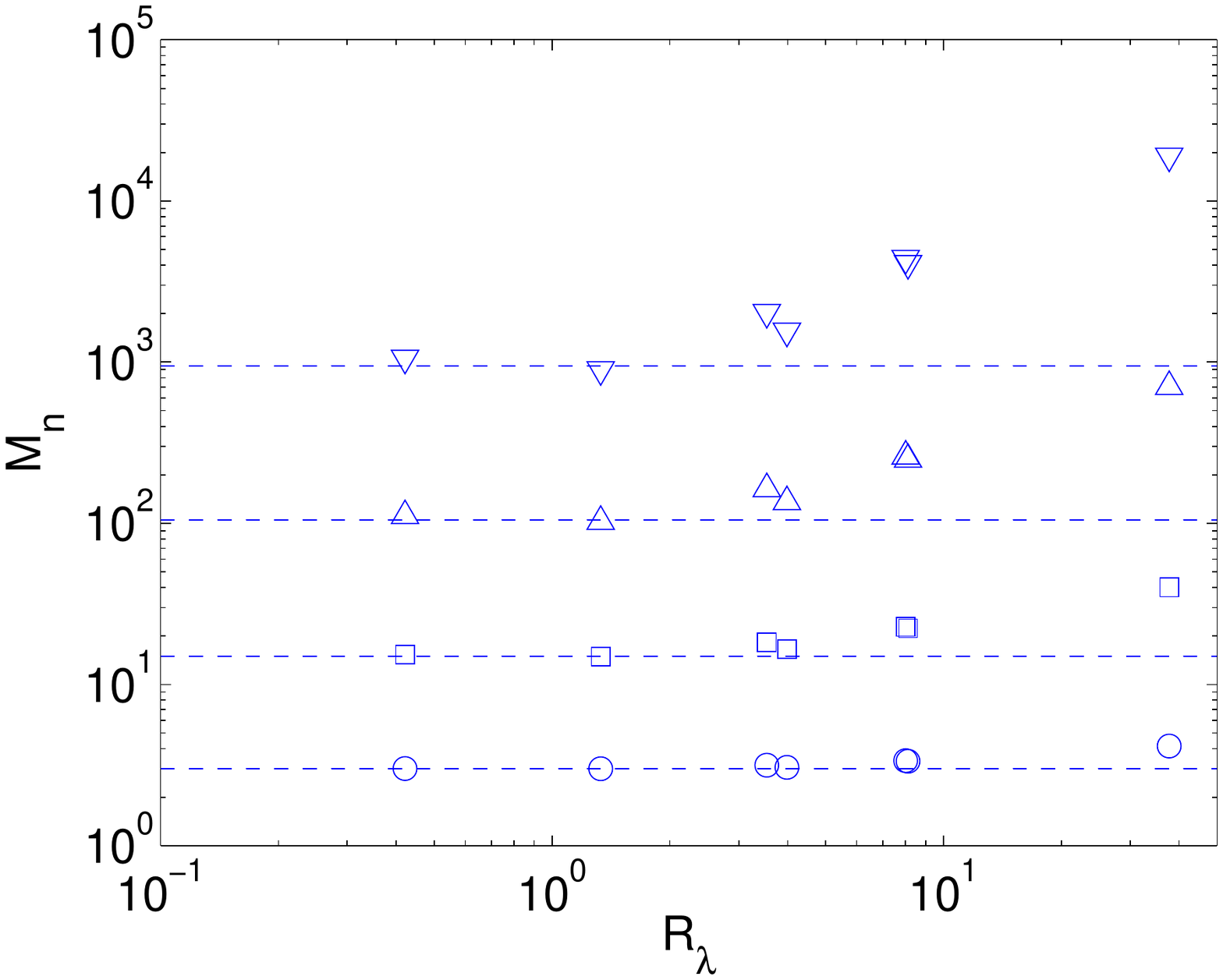}
\includegraphics[height=10cm]{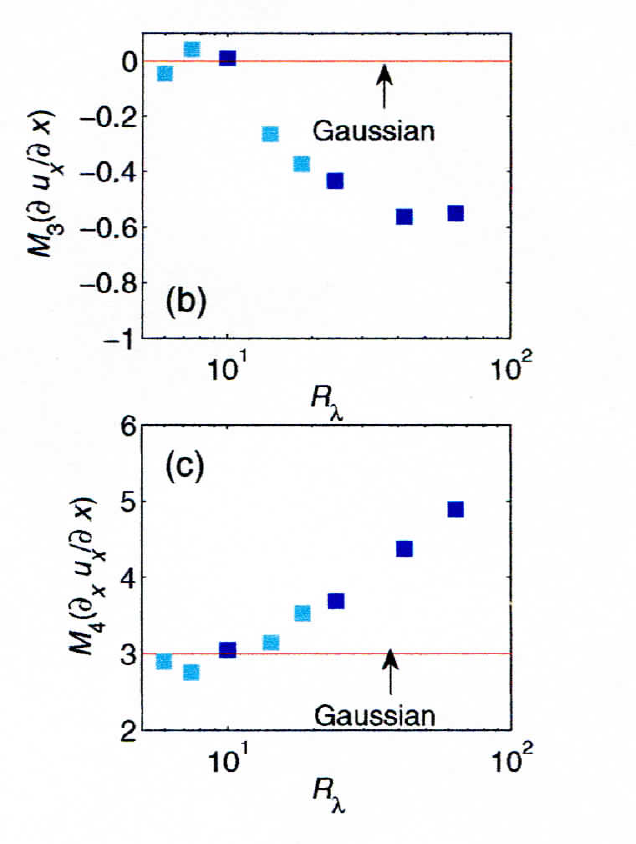}
\caption{(Color online) Normalized moments of velocity derivatives  in Isotropic and Homogeneous turbulence.  Numerical simulations [3]-[4].  Right:  Schumacher et al.[ 3]: homogeneous turbulence driven by  a force ${\bf F}({\bf k},t)={\cal P}\frac{{\bf u}({\bf k},t)}{\sum' |\bf {u}({\bf k},t)|^{2}}\delta_{\bf k,k'}$ (for more details, see above). Left: results of Donzis, Yeung \& Sreenivasan  [4]: HIT drven by a gaussian large-scale noise.
In the range $R_{\lambda}\leq 9.0$, we see a clear Gaussian behavior of a few moments with $S_{3}=0$; $M_{4}\approx 3$; $M_{6}\approx 15$; $M_{8}\approx 105$, typical of Gaussian distribution. At $R_{\lambda}\geq 9-10$, all  moments obey  anomalous scaling of fully developed turbulence}
\label{figure 3}
\end{figure*}

\noindent On Fig.3  the moments $M_{n}$ of velocity derivatives  are shown in the vicinity of a transition point.  One can see that at $R_{\lambda}=9-10$ a relatively sharp transformation from a sub-Gaussian at $R_{\lambda}<9-10$ to anomalous scaling of the dissipation rate moments occurs independently on the driving force. This surprising result will be used below as a constraint on development of turbulence models.\\

\noindent{  {\bf The model.}

Based on  these  results, we consider a flow generated by the Navier-Stokes equations with  a  force ${\bf F}(\Lambda_{f})$. Keeping the force ${\bf F}=const$ and the length-scale $L=2\pi/\Lambda_{f}=const$,  let us  vary viscosity in the interval $ 0\leq \nu \leq \infty$. In the range $\nu>\nu_{tr}$ or $Re<Re_{tr}$ the flow is laminar in a sense that it is described by  a relatively small number of modes with $u({\bf k})$ with $k\approx \Lambda_{f}$. 
\noindent At the transition point  $Re_{\lambda,tr}\approx 9-10$ the transitional pattern ${\bf u}_{0}
(\Lambda_{f})$ is formed, so that:

$$L({\bf u}_{0}, \nu_{tr})=\frac{D{\bf u_{0}}}{Dt}+\nabla p-\nu_{tr}\nabla^{2}{\bf u_{0}}- {\bf F}(\Lambda_{f})\equiv 0$$

\noindent  Here $\frac{D}{Dt}\equiv \frac{\partial}{\partial t}+{\bf v\cdot \nabla}$. When $\nu\ll \nu_{tr}$  the Navier-Stokes equations read:

$$\frac{D{\bf u}}{Dt}=-\nabla p+\nu\nabla^{2}{\bf u}+{\bf F}(\Lambda_{f})\equiv 0$$
  
\noindent  If we write ${\bf u=u_{0}+v}$,  the equation for the ``turbulent'' component ($k>\Lambda_{f}$) of velocity field:

\begin{equation} 
\frac{\partial {\bf v}}{\partial t}+{\bf v}\cdot\nabla {\bf v}=-\nabla p+\nu\nabla^{2}{\bf v}+{\bf f }
\end{equation}


\noindent with 

$${\bf f}={\bf f_{1}+f_{2}+f_{3}}=-{\bf u_{0}\cdot \nabla v}-{\bf v\cdot \nabla u_{0}}+(\nu-\nu_{tr})\nabla^{2}{\bf u}_{0}(\Lambda_{f})$$. 

\noindent The first term ${\bf f}_{1}$ in this expression describes kinematic transfer of small ``eddies'' by the large ones.  The second, ${\bf f}_{2}$, is responsible for  turbulence production due to  interaction of small-scale fluctuations with the large-quasicoherent flow ${\bf u}_{0}$. This  effect is well-known in the turbulence modeling literature.  

\noindent Thus, the total energy  production rate is :

$${\cal P}=\overline{{\bf f_{2}\cdot v}}=-\overline{v_{i}v_{j}}\frac{\partial  u_{0,i}}{\partial x_{j}}\approx \nu_{T}\overline{(u_{0,i}^{j})^{2}}\approx \Lambda_{f}^{-\frac{4}{3}}\overline{(u_{0,i}^{j})^{2}}$$

\noindent     The balance can be written for each scale $l=2\pi/\Lambda(r)$  but with `` turbulent viscosity'' 
$\nu(r) \equiv \nu(l)\propto {\cal E}^{\frac{1}{3}}l^{\frac{4}{3}}$ and \
introducing the projection operator ${\cal P}_{lmn}=k_{m}P_{ln}({\bf k})+k_{n}P_{lm}({\bf k})$ with  $P_{i,j}=(\delta_{ij}-\frac{k_{i}k_{j}}{k^{2}})$ we have with $\hat{k}=({\bf k},\omega)$:

$$f_{2,l}=-\frac{i}{2}{\cal P}_{lmn}({\bf k})\int v(\hat{q})u_{0,n}(\hat{k}-\hat{q})d\hat{q}$$

\noindent so that $\overline{{\bf f}_{2}}=0$ and

$${\cal P}({\bf k})\propto k^{2}\int \frac{q^{-\frac{13}{3}}}{\Omega^{2}+q^{\frac{4}{3}}}(u^{j}_{0,i})^{2}\delta(\hat{k}-\hat{q})\delta(\omega-\Omega)d\hat{q}d\omega\propto k^{-3}$$

\noindent {\bf The $v$-fluctuations are driven by a pumping  with algebraic spectrum!}

\noindent  The experimental data of Refs.[3]-[6] point to independence of small-scale features of turbulence on the nature 
of production mechanism. On the Fig.2  the energy spectrum  in a flow past circular cylinder  of diameter $D$ is shown for the large - scale Reynolds number  $Re=UD/\nu\approx 10^{6}-10^{7}$.  The  onset of Kolmogorov's  inertial range can be clearly seen  at the wave number $k\approx \Lambda_{f}=2\pi/L$  separating inertial and non-universal, geometry-dependent energy - containing range of scales.  A somewhat striking feature of the plot is a very narrow intermediate range   which points to  the smallness of subleading contributions to the inertial range scaling of the energy  spectrum.
Therefore, for $\nu<\nu_{tr}$ we choose the well-known and well-studied model (1):
 where the  random force,  mimicking small-scale fluctuations  is defined by the correlation function:

\begin{equation}
\overline{f_{i}({\bf k},\omega)f_{j}({\bf k'},\omega')}=2D_{0}(2\pi)^{d+1}k^{-y}P_{ij}({\bf k})\delta({\bf k+k'})\delta(\omega+\omega');\end{equation}

\noindent  Based on the above argument,  the  force (2) is {\it extrapolated'' }  onto    interval  $\Lambda_{f} < k\leq  \Lambda\equiv \Lambda_{0}$, so that $\overline{f_{i}F_{j}}=0$ and, by construction,  $f_{i}(k\leq \Lambda_{f},t)=0$.

 \noindent  
 {\bf The renormalization or coarse graining.} The renormalization group for fluid flows has been developed in Refs.[7]-[8 ] and was generalized to enable computations of various dimensionless amplitudes in the low order in the $\epsilon$-expansion in Refs.[9]-[12]. 
 Introducing velocity and length-and -time  scales  $U=\sqrt{D_{0}/(\nu_{0}\Lambda_{0}^{2})}$,   $X=1/\Lambda_{0}$ and $T
 = t\nu_{0}\Lambda_{0}^{2}$, 
 respectively, the equation (1) can be written as (for simplicity we do not change notations  for dimensionless variables): 

$$
\frac{\partial {\bf v}}{\partial T}+\hat{\lambda}_{0}{\bf v\cdot \nabla v}=-\hat{\lambda}_{0}{\nabla}p +\nabla^{2}{\bf v}+ \frac{{\bf f}}{\sqrt{D_{0}\nu_{0}\Lambda^{2}}}
$$

\noindent where the single dimensionless coupling constant (``bare'' Reynolds number) is:  $\hat{\lambda_{0}}^{2}=\frac{D_{0}}{\nu_{0}^{3}\Lambda_{0}^{\epsilon}}$

\noindent{\it Projecting Navier-Stokes equation onto domain $k\leq \Lambda_{0}e^{-r}$ where $r\rightarrow 0$}. 
  {\bf Technical  details of  all calculations presented below are best described in [10]. }
Formally introducing   modes $v^{<}({\bf k},t)$ and $v^{>}({\bf k},t)$ with $k$ from the intervals $k\leq \Lambda_{0} e^{-r}$  and 
$\Lambda_{0}^{-r}\leq k\leq \Lambda_{0}$, respectively,  and averaging  over small-scale fluctuations ${\bf v}^{>}$, leads to equation for the large-scale modes:

\begin{equation}
\frac{\partial v_{i}^{<}}{\partial t}+ v_{j}^{<}\cdot \nabla_{j} v_{i}^{<}=-\nabla_{i} p +\frac{\partial {\sigma_{ij}^{(2)}}}{\partial x_{j}}+
\nu\nabla^{2} v_{i}^{<}+f_{i}+ \Delta f_{i} 
\end{equation}

\noindent  where the second - order correction to the Reynolds stress $\sigma_{ij}=-\overline{v_{i}v_{j}}$ is:

\begin{eqnarray}  
\sigma^{(2)}_{ij}=\hat{\lambda}^{2}(r)\Delta\nu(r)S_{ij}-\hat{\lambda}^{4}_{1}(r)\nu(r)\frac{D}{Dt}[\tau(r)S_{ij}]-\nonumber\\
\hat{\lambda}^{4}_{1}(r)\nu(r)\tau(r)[\beta_{1}\frac{\partial u_{i}}{\partial x_{k}}\frac{\partial u_{j}}{\partial x_{k}}          +\beta_{2}\big(\frac{\partial u_{i}}{\partial x_{k}}\frac{\partial u_{k}}{\partial x_{j}} +
\frac{\partial u_{j}}{\partial x_{k}}\frac{\partial u_{k}}{\partial x_{i}}\big) \nonumber\\
+\beta_{3}\frac{\partial u_{k}}{\partial x_{i}}\frac{\partial u_{k}}{\partial x_{j}} ] 
+O(\hat{\lambda}_{1}^{6})+\cdot\cdot\cdot
\end{eqnarray}

\noindent where $\hat{\lambda}_{1}=O(\hat{\lambda}_{0}(e^{-r}-1))$ is a coupling   constant generated by the scale-elimination and the time-constant $1/\tau(r)=\nu(r)\Lambda^{2}(r)$.  In this limit the coefficients $\beta_{i}$ can be explicitly calculated  [13]  The ``dressed''  viscosity is  denoted as $\nu(r)=\nu+\Delta\nu(r)$ with 
 correction to ``viscosity''   written in the wave-number space as:
\begin{equation}
\Delta\nu=A_{d}\frac{D_{0}}{\nu_{0}^{2}}[\frac{e^{\epsilon r}-1}{\epsilon \Lambda^{\epsilon}_{0}}+O(\frac{k^{2}}{\Lambda_{0}^{\epsilon+2}}\frac{e^{(\epsilon+2)r}-1}{\epsilon +2}) +O(\hat{\lambda}_{0}^{4})]
\end{equation}
 
\noindent   $\epsilon=4+y-d$ and 
$A_{d}=\hat{A_{d}}\frac{S_{d}}{(2\pi)^{d}}; \  \hat{A_{d}}=  \frac{1}{2}\frac{d^{2}-d}{d(d+2)}$.  On  the interval $k<\Lambda_{0}e^{-r}$ the equations (3)-(5) are equivalent to the original equations of motion defined on the interval $k\leq \Lambda_{0}$..

\noindent {\it Iterating  scale-elimination procedure.} The relations (3)-(5) are exact as long as the eliminated ``slice'' in the wave-vector space is very thin.  As will be shown below, in the limit $\Lambda(r)\ll \Lambda_{0}$ the high-order contributions to  the Reynolds stress are not small. The problem is that due to proliferation of tensorial indexes , these terms, while can be qualitatively analyzed using  Wyld's  diagrammatic expansion [14],  are very hard to calculate.\\  

\noindent Eliminating the modes from the interval $\Lambda_{0}e^{-r}\leq k\leq \Lambda_{0}$ the equations can be formally written:

\begin{equation}
\frac{\partial v_{i}^{<}}{\partial t}+ v_{j}^{<}\cdot \nabla_{j} v_{i}^{<}=-\nabla_{i} p +
(\nu+\Delta\nu)\nabla^{2} v_{i}^{<}+{\bf f }+ \Delta f_{i}  +HOT
\end{equation}

\noindent  where,  by  Galileo invariance,  high-order  $(n>1)$  contributions generated by the scale-elimination  can be formally written:

\begin{eqnarray}
HOT=[\sum_{n=2}^{\infty}\hat{\lambda_{1}}^{2n}(r)\tau^{n-1}(r)(\partial_{t}{\bf v^{<}}+{\bf v^{<}\cdot \nabla)^{n}]v^{<}} +\nonumber \\
O(\hat{\lambda}_{2}(r)^{4}\nabla S^{2}_{ij}\frac{1}{\Lambda^{2}(r)}\frac{e^{(\epsilon+2)r}-1}{\epsilon+2}) +\cdot\cdot\cdot
\end{eqnarray}

\noindent with   $\tau(r)\approx 1/(\nu(r)\Lambda^{2}(r)$ and $\hat{\lambda}_{1}=\hat{\lambda}_{0}(e^{\epsilon r}-1)$.  In addition, the expressions  (6)-(7) include various products of time- and space-derivatives responsible, for example, for the rapid distortion effects (RDE).  The high-order nonlinearities generated by the procedure are small if the eliminated shell is very thin but, as will be shown below, they exponentially grow with increase of    $r$. It is clear that the procedure generates an infinite number of coupling constants which are the factors in front of non-linear terms.



\noindent {\it  Recursion relations.}     As a result of  elimination of the first shell $\Lambda_{0}e^{-r}\leq k \leq \Lambda_{0}$,  the original uncorrected ``bare'' viscosity $\nu_{0}$ disappears and instead the equations include only ``dressed''  viscosity $\nu(r)$. Then, starting with the equations  (4)-(7) defined on the   interval $k<\Lambda_{o}e^{-r}$, we can eliminate the  modes from the next shell of wave-numbers  $\Lambda_{0}e^{-(r+\delta r)}\leq k\leq \Lambda_{0}e^{-r}$ and derive  equations  of motion with another  set of corrected  transport coefficients.  The procedure can be iterated resulting in  the cut-off-dependent  viscosity, induced force etc.   Therefore, with $\delta r\rightarrow 0$ the parameters in  the coarse-grained ``Navier-Stokes equations''   for the ``resolved'' velocity field ${\bf u}^{<}$,  defined at  the scales $l\geq 2\pi/\Lambda(r)$,  satisfy the differential relations:

\begin{eqnarray}
\frac{\nu(r+\delta r)-\nu(r)}{\delta r}=A_{d}\frac{D_{0}}{\nu(r)^{2}}\frac{1}{ \Lambda^{\epsilon}(r)}\big{[}\sum_{n=0}^{\infty}\alpha_{n}\hat{\lambda}^{n}(r)+  \nonumber \\
O(\frac{k^{2}}{\Lambda(r)^{2}} )\big{]}
\end{eqnarray}
 
\noindent and 

\begin{eqnarray}
HOT=[\sum_{n=2}^{\infty}\hat{\lambda_{1}(r)}^{2n}\tau(r)^{n-1}(\partial_{t}{\bf u^{<}}+{\bf u^{<}\cdot \nabla)^{n}]u^{<}} +\nonumber \\
O(\hat{\lambda}_{2}(r)^{4}\nabla S^{2}_{ij}\frac{1}{\Lambda^{2}(r)}) +\cdot\cdot\cdot
\end{eqnarray}

\noindent with   $\tau(r)\approx 1/(\nu(r)\Lambda^{2}(r))$.   To asses the role of  different contributions to (8)-(9), first we {\bf assume}   $\hat{\lambda}(r)\ll1 $ and  
$\hat{\lambda}_{1}(r)\ll 1$ and analyze the lowest-order terms  only.\\

\noindent  {\bf Low-order  truncation of the expansion (8)-(9). } This leads  to differential recursion  equations:  recalling that  $\Lambda(r)=\Lambda_{0} e^{-r}$, one obtains:

$$
\frac{\nu(r+\delta r)-\nu(r)}{\delta r}=\frac{d\nu(r)}{d r}=A_{d}\nu(r)\hat{\lambda}^{2}(r)  
$$

$$\frac{d\hat{\lambda}^{2}}{dr}=\epsilon\hat{\lambda}^{2}-3A_{d}\hat{\lambda}^{4}$$

\noindent where $\hat{\lambda}^{2}(r)=\frac{D_{0}}{\nu^{3}(r)\Lambda^{\epsilon}(r)}$ 
and: 

$$ \nu(r)=
 \nu_{0}[1+\frac{3A_{d}}{\epsilon \nu_{0}^{3}}\frac{D_{0}S_{d}}{(2\pi)^{d}}(\frac{1}{\Lambda^{\epsilon}(r)}-\frac{1}{\Lambda_{0}^{\epsilon}})]^{\frac{1}{3}}$$

 $$\hat{\lambda}
= \hat{\lambda}_{0}e^{\frac{\epsilon r}{2}}[1+\frac{3A_{d}}{\epsilon \nu_{0}^{3}}\frac{D_{0} S_{d}}{(2\pi)^{d}}(\frac{1}{\Lambda^{\epsilon}(r)}-\frac{1}{\Lambda_{0}^{\epsilon}})]^{-\frac{1}{2}}
$$
 
 \noindent The solution for the ``induced''  coupling  constant $\hat{\lambda}_{1}$ is :
 
 \begin{equation}
 \hat{\lambda}_{1}(r)=\frac{\sqrt{\epsilon}e^{\frac{\epsilon r}{2}}}{\sqrt{\frac{\epsilon}{\hat{\lambda}_{1}^{2}(0)}+3A_{d}(e^{\frac{\epsilon r}{2}}-1)}}
 \end{equation}

\noindent    For $\epsilon=4$, corresponding to Kolmogorov's energy spectrum,  in the limit $\epsilon r\gg1$ the coupling constants tend to the fixed point

\begin{equation}
\hat{\lambda}_{*}\rightarrow (\frac{\epsilon}{3A_{d}})^{\frac{1}{2}} \approx 1.29\sqrt{\epsilon}\approx 2.58. 
\end{equation} 
 
\noindent   It is also clear that $\hat{\lambda}_{1,*}\approx  \hat{\lambda}_{*}$.  {\it This result means that for $k>\Lambda_{f}$, the above truncation of the expansion is,  in general,  incorrect  and high-order non-linearities generated by procedure are not small. Now we consider a special case of the transport approximation $\Lambda(r)\rightarrow \Lambda_{f}=2\pi/L$.}\\

\noindent {\it Parameters.} All low-order calculations leading to dimensionless amplitudes  presented below,  are best described in great  detail in Ref.[10].  Eliminating all modes from the interval $k\geq \Lambda_{0}e^{-r}$ and setting $\epsilon=4$  gives:

\begin{eqnarray}
\nu(k)=(\frac{3}{8}A_{d}2D_{0})^{\frac{1}{3}}k^{-\frac{4}{3}}\approx 0.42(\frac{2D_{0}S_{d}}{(2\pi)^{d}})^{\frac{1}{3}}k^{-\frac{4}{3}}\nonumber
\end{eqnarray}

\noindent and from the linearized  equation   at the fixed point
 we derive Kolmogorov's spectrum valid in the range $k\geq\Lambda_{f}$ :

\begin{eqnarray}
E(k)=\frac{1}{2}\frac{S_{d}k^{2}}{(2\pi)^{d+1}}\int_{-\infty}^{\infty}TrV_{ij}({\bf k}\omega)d\omega=\nonumber \\
\frac{1}{2(\frac{3}{8}\hat{A}_{d})^{\frac{1}{3}}}(2D\frac{S_{d}}{(2\pi)^{d}})^{\frac{2}{3}}k^{-\frac{5}{3}} =\nonumber \\
1.186(2D_{0}\frac{S_{d}}{(2\pi)^{d}})^{\frac{2}{3}}k^{-\frac{5}{3}}
\end{eqnarray}

\noindent where $(2\pi)^{d+1}V_{ij}({\bf k},\omega)=\frac{u^{<}_{i}({\bf k},\omega)u^{<}_{j}({\bf k'},\omega')}{\delta{\bf k+k'})\delta(\omega+\omega')}$.  In the so called EDQNM  approximation, which is exact at the Gaussian fixed point (see below),    the force amplitude $D_{0}$ can be related to the  mean dissipation rate [10], [11]:

\begin{equation}
2D_{0}S_{d}/(2\pi)^{d}\approx 1.59{\cal E};  \ E(k)=C_{K}{\cal E}^{\frac{2}{3}}k^{-\frac{5}{3}};  \ C_{K}=1.61
\end{equation}

\noindent  Let us identify  the  infra-red cut - off $\Lambda_{f}=\Lambda(r)\approx 2\pi /L$  with the wave-number corresponding to  the top of the inertial range. In the large Re-limit $\Lambda_{0}/\Lambda_{f}\gg1$, the total energy of the inertial range turbulent fluctuations is evaluated readily:

\begin{eqnarray}
{\cal K}=\int_{\Lambda_{f}}^\infty E(k)dk=\frac{3}{2}C_{K}(\frac{{\cal E}}{\Lambda_{f}})^{\frac{2}{3}}=\nonumber \\
\frac{3}{2}1.61(\frac{3}{8}\hat{A}_{d}1.59)^{\frac{1}{3}}\frac{{\cal  E}}{\nu(\Lambda_{f})\Lambda_{f}^{2}}\approx 
1.19\frac{{\cal  E}}{\nu(\Lambda_{f})\Lambda_{f}^{2}}
\end{eqnarray}

\noindent and,  setting $k=\Lambda_{f}$   gives the  expression for effective viscosity in  equation for the large-scale dynamics in the interval of scales $k<\Lambda_{f}$:

\begin{equation}
\nu_{T}\equiv  \nu(\Lambda_{f})\approx 0.084\frac{{\cal K}^{2}}{{\cal E}};  \hspace{1cm} 10.0\times \nu(\Lambda_{f})^{2}\Lambda_{f}^{2}={\cal K}
\end{equation}\\








 
{ \bf Fixed- point Reynolds number and irrelevant variables.}    The expression (15) gives effective viscosity accounting for all turbulent fluctuations from the interval $1/\Lambda_{0}\leq r<L=1/\Lambda_{f}$ acting on  the almost-coherent-large scale flow on the scales $r\approx L=1/\Lambda_{f}$.  Using (13) -(15)  we can calculate the effective 
$R_{\lambda,f}={2\cal K}\sqrt{5/(3{\cal E}\nu(\Lambda_{f}))}=\sqrt{20/(3\times 0.084)} =9.0$. 
The same parameter   can be expressed in terms of the fixed-point coupling constant:

 

\begin{eqnarray} 
\hat{\lambda}^{*}=\sqrt{\frac{D_{0}S_{d}/(2\pi)^{d}}{\nu_{T}^{3}\Lambda_{f}^{4}}}=
\sqrt{\frac{0.8{\cal E}}{\nu_{T}^{3}\Lambda_{f}^{4}}}=\frac{\sqrt{0.8\times 400{\cal E}\nu_{T}}}{u_{rms}^{2}}\nonumber \\
=\frac{\sqrt{0.8\times 400\times \frac{5}{3}}}{R_{\lambda}^{fp}}=\sqrt{\frac{4}{3\hat{A}_{d}}} =2.58\nonumber
\end{eqnarray}

\noindent and  $R_{\lambda,f}\approx 9.0$ very close to  Reynolds number of transition $R_{\lambda}\approx 9.$,   
obtained from direct numerical simulations of Refs.[3]-[4].   This result agrees   with observation that in the flows past various bluff bodies, 
the Reynolds number based on the measured ``turbulent viscosity''  and large-scale velocity  field is $R_{\lambda,T}=O(10)$,  independent on the ``bare''  (classic) Reynolds number calculated with molecular viscosity.  


As $\Lambda(r)\rightarrow L=2\pi/\Lambda_{f}$,  the effective viscosity  $\nu(r)\rightarrow \nu_{tr}$  and $Re(r)\rightarrow Re_{tr}$, which is the most important and surprising outcomes of the theory.    If, as was found numerically,  transition to turbulence is ``smooth'',  
the velocity field ${\bf u}_{0}$ must come out from   equations of motion obtained by the scale -elimination and  Navier-Stokes equations for quasi-laminar flow at a transition point: therefore

\begin{equation}
\hat{L}({\bf u}_{0},\nu_{\Lambda_{f}})-HOT=\hat{L}({\bf \bf u_{0}},\nu_{tr})\equiv 0
\end{equation}

\noindent If, in addition, the  transition is  universal, i.e. is independent on initial conditions we conclude that as $\Lambda(r)\rightarrow \Lambda_{f}$,  and, according to the above derivation $\nu(\Lambda_{f})\rightarrow \nu_{tr}$, the nonlinearities generated by the scale elimination procedure 

$$HOT\rightarrow 0$$

\noindent are irrelevant. The role of the induced noise will be discussed in detail below.




  
\noindent All we can definitely say  is: if indeed transition is continuos in the limit   $\Lambda(r)\rightarrow \Lambda_{f}$, the nonlinearities $HOT(\Lambda(r))\rightarrow 0$. To understand the way it tends to zero,
let us consider  the linearized equation of motion in the vicinity of the fixed point where   ${\bf u}={\bf u_{0}}+{\bf u}_{1}$:

\begin{equation}
\frac{\partial {\bf u}_{1}}{\partial t}+{\bf u}_{0}\cdot \nabla {\bf u}_{1}+{\bf u}\cdot \nabla {\bf u}_{0}=-\nabla p_{1}+\nu\nabla^{2}{\bf u}_{1}+HOT
\end{equation} 

\noindent  If ${\bf u}_{1}\propto Ae^{i\omega t}$, then according to purely phenomenological Landau's theory of transition:  $u_{1}\propto  A_{max}\propto \sqrt{Re-Re_{tr}}$


\noindent  and 

$$HOT\approx{\bf u}_{0}\cdot \nabla {\bf u}_{1} \approx u^{2}_{0}\Lambda_{f}\sqrt{Re-Re_{tr}}$$


{\it Large-scale dynamics.} Now we would like to discuss the large-scale flow in the interval $k\approx \Lambda_{f}$, where  the bare force ${\bf f}({\bf k})=0$ and therefore the equation of motion is:

\begin{eqnarray}
\frac{\partial {\bf v^{<}}}{\partial t}+{\bf v^{<}\cdot \nabla v^{<}}=-{\nabla}p +
 \nu(\Lambda_{f})\nabla^{2}{\bf v^{<}}+ {F+\bf \psi}
\end{eqnarray}

\noindent with induced noise evaluated in Ref.[10]:

\begin{equation}
\overline{\psi_{i}({\bf k})\omega)\psi_{j}({\bf k'},\omega')}= (2\pi)^{d+1}2D_{L}k^{2}P_{ij}({\bf k})\delta({\bf k+k'})\delta(\omega+\omega')
\end{equation}

\noindent where 

\begin{equation}
D_{L}=D_{0}\frac{d^{2}-2}{20d(d+2)}\frac{\hat{\lambda}^{*}\hat{\lambda}^{*}}{\Lambda_{f}^{5}}=D_{0}\frac{0.155}{\Lambda_{f}^{5}}
\end{equation}

\noindent The induced force $\psi$ is the result of  small-scale turbulent fluctuations on the large-scale dynamics which is often called ``backscattering''. 
In the most important limit $k\rightarrow \Lambda_{f}$  the ratio
$
D_{0}/D_{L}\approx 1/0.155 \approx 7.0$, which, while being  numerically   not too  large,  is responsible for both 
blurring of the  large-scale transitional patterns and for the observed   
gaussian statistics  of the large-scale velocity fluctuations  in the high-Reynolds number flows . \\


 \noindent {\bf Summary and conclusions.}   The theory of critical phenomena  is  one of the most spectacular  
 applications of  renormalization group to statistical mechanics. In this approach the procedure is  not applied to microscopic Hamiltonians, like those for ferromagnetic solids  or  liquid helium,  but to macroscopic Hamiltonians (free energies) reflecting basic large - scale symmetries  of a  system.   The magnitude  of  critical temperature $T_{c}$,  depending on the details of intermolecular microscopic interactions remains undetermined and the size of the system does not appear in the theory where  all correlation functions are expressed in terms of $\tau=\frac{|T-T_{c}|} {T_{c}}$. 
 In this respect, hydrodynamics are  different, for large-and small-scale dynamics 
 responsible for  transition at  $Re=Re_{tr}$  and  behavior   of turbulent fluctuations  at $Re>>Re_{tr}$,   are contained in the Navier-Stokes equations.  Therefore, it is  not surprising that   the recently discovered universal $Re_{\lambda,tr}\approx 9.-10$ can serve as a dynamic constraint on the theory. 

  1. \  One of most difficult, not yet understood properties of turbulence,  is anomalous scaling of velocity increments    on the small scales $\eta\ll l  \ll  L=2\pi/\Lambda_{f}$.  It was shown above that in this range of scales the  flow is described by an infinite number of  O(1) coupling constants.  At this point we do not know how to deal with them.  

2.  The selection of relevant variables is possible in the limit $l \rightarrow \L$.  Keeping only  the lowest -order contributions,   the calculated fixed-point Reynolds number $Re_{f}\approx Re_{\lambda,tr}$ where $Re_{\lambda,tr}=9.-10$ is  the  numerically computed Reynolds number of transition to turbulence.  Since  the numerically discovered transition is ``smooth'',   i.e.  ${\bf u}_{0}={\bf u}_{fp}$ and  ($\nabla_{i}u_{0,j}=\nabla_{i}u_{fp,j}$),   at this point all additional to the NS equations high-order nonlinearities are irrelevant.   \\

3. \   Comparison with Landau's  theory of transition to turbulence shows that in the vicinity of transition point, the neglected nonlinear terms are $O(\sqrt{Re-Re_{tr}})\rightarrow 0$.\\
 
4. The infra-red divergencies appearing in the each term of the expansion do not disappear but are summed up into equations of motion for the large-scale features of the flow. 
\noindent To stress how accurate the derived transport approximation is, we would like to reproduce our old result  on decay of isotropic and homogeneous turbulence [10].  Since in this flow $\overline{S}_{i,j}=0$, the equations governing the decay are very simple:

$$\frac{\partial {\cal K}}{\partial t}=-{\cal E};\hspace{1cm} \frac{\partial {\cal E}}{\partial t}=-C_{\epsilon,2}\frac{{\cal E}^{2}}{{\cal K}}$$

\noindent with $C_{\epsilon,2}=1.68$ calculated at the integral scale in the lowest order of renormalized perturbation expansion [10].   The present paper justifies the approximation and procedure  leading to this and all other constants calculated in [10] by an ad hoc  neglecting of HOT. 
The above  equations give:

$${\cal K}/{\cal K}_{0}=(\frac{t}{t_{0}})^{-\frac{1}{C_{\epsilon,2}-1}}=(\frac{t}{t_{0}})^{-\gamma}\approx (\frac{t}{t_{0}})^{-1.47}$$

\noindent This result has a long and difficult history. First, it was shown by Kolmogorov that $\gamma=10/7\approx 1.43$, very close to the one shown above.  Somewhat later,  Kolmogorov's construction has been   reinterpreted by Landau as a consequence of conservation of the angular momentum  [2].  
This theory has been criticized by Batchelor et. al.  [15] and the early experiments,   yielding   $\gamma\approx 1.0-1.3$ (see  for example Ref.[ 16]),  seemed  to support Batchelor's conclusions.   A huge number of experimental,  theoretical and later numerical papers  dealt with this subject [16].  As a consequence,   the constant $C_{\epsilon,2}$ in the ${\cal K}-{\cal E}$ model,  widely  used for engineering simulations, was taken as $C_{\epsilon,2}\approx 1.92$  corresponding $\gamma\approx 1.1$.  This led  to the over-dissipated  turbulent velocity field  computed with this model.   It took  many years to realize that Kolmogorov's theory was developed for a  finite patch  of turbulence in  an infinite fluid  and the exponent $\gamma$ was very sensitive to the finite size effects, geometry  etc. This long-standing confusion has recently been resolved by a remarkable ($4096^{3}$)  numerical simulation  by Ishida et. al.  [17],  who showed that when  the initially prepared flow satisfied constraints of Kolmogorov's theory,  the exponent of kinetic energy decay was  indeed $\gamma\approx 10/7$. \\

5. To conclude we would like to mention that  if,  in general, a  turbulent flow is generated by an   instability of a large-scale quasi-coherent flow pattern (dynamical system) ${\bf u}_{0}$, then the equations of motion governing anomalous velocity fluctuations are given by (1) with  ${\bf f}_{2}={\bf v\cdot \nabla u_{0}}$. This may explain a broad universality of  small-scale   features of strong turbulence discovered in Refs.[.3]-[6]. 

\noindent I  am grateful to A. M. Polyakov, N.Goldenfeld, V.Lebedev, I Kolokolov, Y. Sinai, U. Frisch, E.Titi , H.Chen,  I. Staroselsky and J.Wanderer  for their interest in this work and  numerous suggestions.  Many ideas leading to this paper emerged from   numerical  investigations  of transition  conducted jointly with   J. Schumacher, D. Donzis and K.R. Sreenivasan.     \\

\noindent {\it References.}\\
 
\noindent 1. K.G. Wilson,  Rev.Mod.Phys. , {\bf 12}, 75 (1974).\\
\noindent 2. \ L.D.Landau \& E.M. Lifshits, ``Fluid Mechanics'', Pergamon, New York, 1982;\\
\noindent 3. \  J.Schumacher, K.R. Sreenivasan \& V. Yakhot; 
New J. of Phys.  {\bf 9}, 89 (2007); \\
\noindent 4. D.A. Donzis, P.K. Yeung and K.R. Sreenivasan, ``Dissipation and enstrophy in homogeneous turbulence: resolution effects and scaling in direct numerical simulations'', Phys.Fluids {\bf 20}, 045108 (2008); \\
\noindent  5. \   P.E. Hamlington, D. Krasnov, T. Boeck and J. Schumacher, ``Local dissipation scales and energy dissipation - moments in channel flow'', J. Fluid. Mech. {\bf 701}, 419-429 (2012);\\ 
\noindent  6. J. Schumacher, J. D. Scheel, D. Krasnov, D. A. Donzis, V. Yakhot and K. R. Sreenivasan,
Proc. Natl. Acad. Sci. USA {\bf 111}, 10961-10965 (2014)
 \\
\noindent  7. \ D. Forster, D. Nelson \&  M.J. Stephen, Phys.Rev.A {\bf 16}, 732 (1977);\\
\noindent 8 \ C. DeDominisis \& P.C. Martin, Phys.Rev.A{\bf 19}, 419 (1979);\\
\noindent  9. \  
 V. Yakhot \&  S.A. Orszag, Phys.Rev.Lett.{\bf 57}, 1722 (1986);\\
\noindent  10.  \ V. Yakhot \& L. Smith, 
J. Sci.Comp. {\bf 7}, 35 (1992);\\
\noindent 11.  \ V. Yakhot, S.A. Orszag, T. Gatski, S. Thangam \& C.Speciale, 
 Phys. Fluids A{\bf 4}, 1510  (1992);\\
 \noindent 12. \ W.P. Dannevik,  V. Yakhot \& S.A.Orszag, Phys.Fluids {\bf 30}, 2021 (1987);\\
  \noindent 13. \ R.Rubinstein \& M.Barton, Phys.Fluids, {\bf A12}, 1472  (1990);  \ 
  H.Chen, S.A. Orszag, I.Staroselsky \& S.Succi, J.Fluid Mech, {\bf 519}, 301 (2004);\\
\noindent 14. \  H.W. Wyld,   
Annals of Physics {\bf 14}, 143-165 (1961);  
 \noindent  15. \ G.K. Batchelor and I. Proudman, Trans. R. So. Lond. A{\bf 248}, 369-405 (1956);\\
\noindent 16 . \  G. Compte-Bellot and S. Corrsin,  ``The use of a contraction to improve the isotropy of grid-generated turbulence'', J.Fluid Mech {\bf 25}, 657-682 (1966); A.S. Monin and  A.M. Yaglom, ``Statistical Fluid Mechanics'', The MIT Press, v2, Cambridge, MA., 1975\\
\noindent 17. T.Ishida, P.A. Davidson and Y.  Kaneda, J.Fluid.Mech, {\bf 564}, 455-475 (2006); \\

\end{document}